# Permanent Magnet System for MRI with Constant Gradient mechanically adjustable in Direction and Strength

Peter Blümler

Institute for Physics, University of Mainz, 55099 Mainz, Germany
email: bluemler@uni-mainz.de

**Abstract:**

A design for a permanent magnet system is proposed that generates spatially homogeneous, constant magnetic field gradients, thus creating conditions suitable for MRI without gradient coils and amplifiers. This is achieved by superimposing a weak Halbach quadrupole on a strong Halbach dipole. Rotation of either the quadrupole or the entire magnet assembly can be used to generate 2D images via filtered back-projection. Additionally, the mutual rotation of two quadrupoles can be used to scale the resulting gradient. If both gradients have identical strength the gradient can even be made to vanish. The concept is demonstrated by analytical considerations and FEM-simulations.

# Introduction:

In recent years several permanent magnet systems were developed for the emerging field of compact or even mobile NMR [1, 2]. Many of them were also equipped with classical gradient coils driven by electrical currents to allow for MRI applications. A particularly suitable magnet design for this purpose is the Halbach cylinder which nowadays can be constructed and shimmed to produce images of very high quality [3, 4] and even high-resolution NMR-spectra [5]. However, shimming permanent magnets is still a demanding iterative process. In 2014 Wald et al. presented a method to produce MR-images utilizing the inhomogeneous magnetic field of a non-shimmed Halbach [6]. This was done by collecting spatial information from exact knowledge of the field, mechanical rotation of the entire magnet, using an eight channel receiver array and image reconstruction via special software. On a first glance this appears like a step back to the very first MRI methodology where images were reconstructed from one-dimensional projections along different angles using filtered Radon-transforms (back-projection). The direction of the projection was changed by either rotating the gradient or the sample [7]. On the other hand, getting rid of the very power consuming gradient amplifiers helps to make MRI simpler, more economical and portable. However, there is a slight problem with the proposed method in [6] because the image reconstruction is limited in regions of rather homogeneous fields (e.g. in the center) this and other ambiguous field regions could only partially be corrected for by using multiple receive coils and channels.

The aim of this contribution is to follow up on this inspiring idea and suggests a design that removes the problems of nonbijective projections. Instead of using an inhomogeneous field with ambiguities it is proposed to use constant gradients of high spatial homogeneity superimposed on a Halbach dipole and hence, allow for scaling their direction and strength, eventually even switch them off, by rotating additional rings of permanent magnets. Like in [6] only permanent magnets are involved, which however generate better defined inhomogeneities that are scalable in amplitude and direction. It is hoped that simple Fourier-transformation and Radon-transformation (filtered backprojection) will be sufficient to generate images. Of course this can also be combined with multiple receivers to speed up acquisition via parallel MRI and non-linear image reconstruction schemes to relax the demands of construction accuracy.

This paper only introduces the concept by calculations and FEM-simulations only, and no such device has been built yet by the author.

## Concept:

Klaus Halbach's original paper [8] already introduces the fundamental idea, because he was particularly interested in producing magnetic fields with many poles for e.g. accelerators. The basic construction principle for this is a cylinder of permanent magnet material where the direction, $\alpha$, of the magnetization is a multiple of its location angle $\theta$ (see Fig. 1a).

$$\alpha = (m+1)\,\theta \quad \text{with} \quad m \in \mathbb{Z} \qquad [1]$$

The polarity of the resulting field is then given by $2m$ and a positive sign directs the field inwards (with no stray field outside, for an infinitely long cylinder) while a negative sign creates a magnetic field of the same polarity outside the cylinder [4]. For NMR purposes only the case $m = +1$ was used so far, because it produces a homogeneous dipolar field inside the cylinder (see Fig. 1b). Halbach ([8] eq. 21) also already provided an analytical expression of the field at position $\vec{r}$ depending on the polarity

$$B = B_R \ln\left(\frac{r_i}{r_o}\right) \qquad \text{for } m = 1$$

$$B(\vec{r}) = \left(\frac{\vec{r}}{r_i}\right)^{m-1} B_R \frac{m}{m-1}\left[1 - \left(\frac{r_i}{r_o}\right)^{m-1}\right] \qquad \text{for } m \geq 2 \qquad [2]$$

where $B_R$ is the remanence of the permanent magnetic material and $r_i$ and $r_o$ are the inner and the outer radius of the cylinder.

The straightforward approach to obtain spatial resolution inside a spatially homogeneous field is to superimpose a much weaker gradient field which transforms the NMR-spectrum into a spatial projection along the gradient direction. It is important to realize that Maxwell's second law ($\vec{\nabla} \cdot \vec{B} = 0$) prevents the generation of a gradient field in one exclusive direction only, and consequently only that component ($dB_y/dr$) of the gradient field is of relevance which aligns with the direction of the much stronger homogeneous field ($B_0$ here along $y$-direction) needed for polarizing the nuclear spins. The other components ought and will not be noticeable[1] due to the limited bandwidth of the NMR excitation. Such a linear field dependence or constant gradient field can either be obtained from a coil or a permanent magnet of quadrupolar polarity.

---

[1] If this is not the case because, for instance in low field MRI, the other (concomitant) gradient component has a similar strength as $B_0$, the full gradient tensor needs to be considered and image reconstruction is no longer possible by simple Fourier-transformation [9]

A quadrupole field (e.g. Fig. 1c) consist of two orthogonal constant gradient fields (see Fig. 1 e and f) while its magnitude has a conical shape (see Fig. 1d).

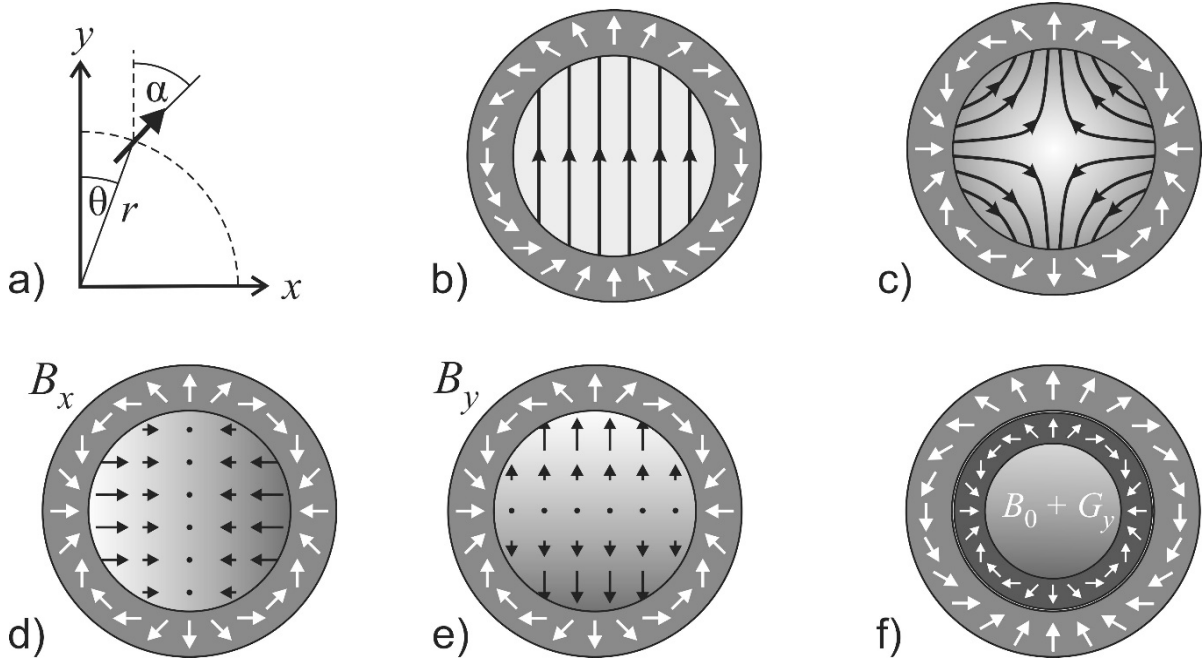

Fig. 1:  Schematic illustration of the magnet geometries: a) Construction principle of Halbach-cylinders of polarity $2m$. Magnetic material placed with its center on a circle (dashed line) at an angle $\theta$ has its magnetization direction (bold arrow) at an angle $\alpha = (m + 1)\,\theta$. The coordinate system is used for all other graphs. b) Inner Halbach dipole ($m = +1$) consisting of a cylinder of permanent magnet material (dark gray) with continuously changing magnetization direction as indicated by the white arrows. The gray shade in the center illustrates the amplitude of the field, while its direction is specified by the black arrowed flux lines. c) Inner Halbach quadrupole ($m = +2$) in the same representation as b). d) and e) show the $B_x$ and $B_y$ component of the flux in c). Here the black arrows indicate strength and direction of the flux (no flux lines!). f) A nested arrangement of b) and c) (in darker gray) producing a superposition of the homogeneous dipole field with a weaker gradient field. Here the quadrupole is inside the dipole, of course it could also encase the dipole.

Figure 1f now shows a dipolar Halbach magnet with its field $B_0$ along the $y$-direction (cf. Fig. 1b in the following indicated by superscript 'D') encasing a smaller quadrupolar Halbach magnet (see Fig. 1c indicated by 'Q'). For such a superposition eq. [2] results in the following vector field inside the geometry

$$\vec{B}(\vec{r}) = \begin{bmatrix} B_x(\vec{r}) \\ B_y(\vec{r}) \end{bmatrix} = \vec{B}_0 + \mathsf{G}\cdot\vec{r} = B_0 \begin{bmatrix} 0 \\ 1 \end{bmatrix} + G \begin{bmatrix} -1 & 0 \\ 0 & 1 \end{bmatrix}\vec{r},$$

$$\text{with }\vec{r} = \begin{bmatrix} x \\ y \end{bmatrix},\quad B_0 = B_R^D \ln\left(\frac{r_i^D}{r_o^D}\right)\quad \text{and}\quad G = 2B_R^Q \left(\frac{1}{r_i^Q} - \frac{1}{r_o^Q}\right) \quad [3]$$

of which only $B_y(x,y) = B_0 + G\,y$ is experimentally relevant (for $B_0 \gg G$). Then MRI experiments can then be performed by rotating either magnet or sample using the backprojection idea presuming that both cylinders are rotated synchronously.

If however the quadrupole is turned by an angle $\beta$ relative to the dipole, the rotation matrix, $\mathsf{R}$, has to be applied to the gradient matrix, $\mathsf{G}$, and $\vec{r}$

$$\vec{B}'(\vec{r}) = \vec{B}_0 + \mathsf{R}\mathsf{G}\cdot(\mathsf{R}^{-1}\vec{r}) \quad \text{with } \mathsf{R} = \begin{pmatrix} \cos\beta & \sin\beta \\ -\sin\beta & \cos\beta \end{pmatrix}$$

$$= B_0 \begin{bmatrix} 0 \\ 1 \end{bmatrix} + G \begin{bmatrix} -x\cos 2\beta + y\sin 2\beta \\ x\sin 2\beta + y\cos 2\beta \end{bmatrix} \quad [4]$$

of which again only the $y$-component is experimentally relevant

$$B'_y(x,y) = B_0 + G(x\sin 2\beta + y\cos 2\beta). \quad [5]$$

The gradient vector of this rotated field, $B'_y$ is then

$$\vec{\nabla} B'_y = \begin{bmatrix} \dfrac{dB'_y}{dx} \\ \dfrac{dB'_y}{dy} \end{bmatrix} = G \begin{bmatrix} \sin 2\beta \\ \cos 2\beta \end{bmatrix}. \quad [6]$$

This means that a rotation of the quadrupole by an angle $\beta$ relative to the dipole causes the resulting field gradient to rotate by $2\beta$ with respect to the direction of $B_0 y$. This is a direct consequence of the $C_2$-symmetry of the quadrupolar field and allows to perform a complete rotation of the gradient direction by turning the quadrupole by only 180° (as illustrated in Fig. 2).

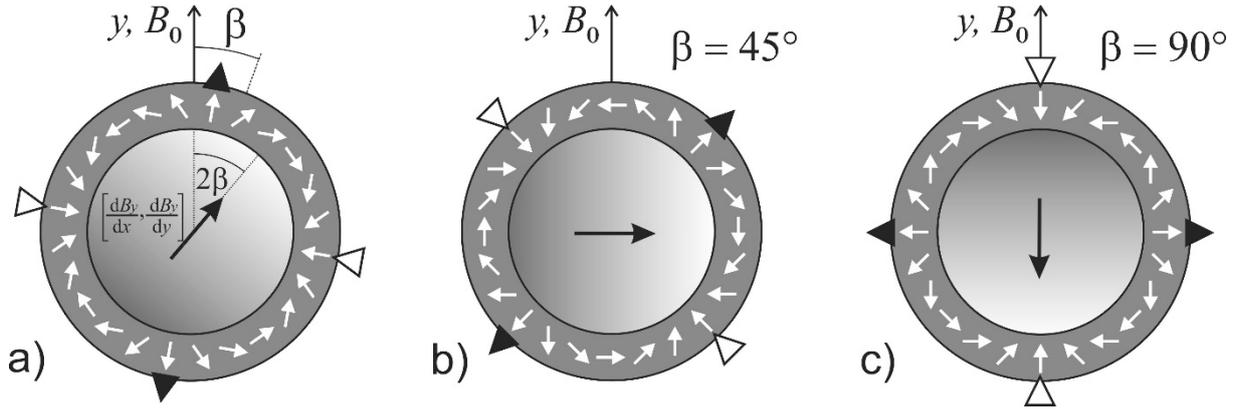

Fig. 2: $B_y$-field component of a Halbach-quadrupole rotated by an angle β in the field of a (Halbach) dipole. For a clearer view the dipole is omitted. Otherwise it would dominate the graph since it should be much stronger than the quadrupole. However, it is necessary because its field direction (here y) selects the field component of the quadrupole, generating a constant gradient [$dB_y/dx$, $dB_y/dy$] which rotates then with 2β. a) Schematic drawing: The original positive y direction of the quadrupole is indicated by a black triangle. The arrow inside the cylinder marks the direction of the gradient whose field change is additionally displayed by a gray scale. b) rotation by β = 45°, c) β = 90°.

While the ability to rotate the direction of the gradient field by turning e.g. a thinner, lighter, outer quadrupole-cylinder might be considered as a minor advantage over rotating the entire magnet, this concept can be extended to change the total strength of the gradient[2]. This is done by placing yet another quadrupole around the structure of dipole and quadrupole. If this second quadrupole is now rotated with respect to dipole and first quadrupole (see Fig. 3a) the gradient field will rotate and change its amplitude. If the first quadrupole is indicated by 'Q1' and the second by 'Q2' their combined field can be expressed by eq. [6] as

$$\vec{G}_\Sigma = \vec{G}_{Q1} + \vec{G}_{Q2} = G_{Q1} \begin{bmatrix} \sin 2\beta \\ \cos 2\beta \end{bmatrix} + G_{Q2} \begin{bmatrix} \sin 2\phi \\ \cos 2\phi \end{bmatrix}$$

$$\text{with} \quad G_{Qi} = 2B_R^{Qi}\left(\frac{1}{r_i^{Qi}} - \frac{1}{r_o^{Qi}}\right) \quad \text{and} \quad i = 1, 2$$

[7]

Where Q1 is rotated by an angle β and Q2 by ϕ relative to $B_0$. If for simplification only Q1 is turned (ϕ = 0) this becomes

---

[2] This is equivalent to changing the total field in amplitude and direction by rotating two or more nested dipoles [10,11]

$$\vec{G}_\Sigma = G_{Q1} \begin{bmatrix} \sin 2\beta \\ \cos 2\beta \end{bmatrix} + G_{Q2} \begin{bmatrix} 0 \\ 1 \end{bmatrix}$$

$$|\vec{G}_\Sigma| = \sqrt{G_{Q1}^2 + 2 G_{Q1} G_{Q2} \cos 2\beta + G_{Q2}^2} \qquad [8]$$

$$\measuredangle \vec{G}_\Sigma = \tan^{-1}\left( \frac{G_{Q1} \sin 2\beta}{G_{Q1} \cos 2\beta + G_{Q2}} \right)$$

Hence, the amplitude of the total gradient field, $|\vec{G}_\Sigma|$, has a maximal value of $|G_{Q1} + G_{Q2}|$ for $\beta = 0°, 180°$ and a minimum of $|G_{Q1} - G_{Q2}|$ at $\beta = 90°, 270°$. In principle the gradient field can be switched off by choosing $G_{Q1} = G_{Q2}$, e.g. by adjusting the outer radius of the second quadrupole to

$$r_o^{Q2} = \frac{B_R^{Q2} r_i^{Q1} r_o^{Q1} r_i^{Q2}}{B_R^{Q1}\left(r_i^{Q1} r_i^{Q2} - r_o^{Q1} r_i^{Q2}\right) - B_R^{Q2} r_i^{Q1} r_o^{Q1}}. \qquad [9]$$

Equations [8] and [9] were tested and confirmed in a 2D-FEM simulation (COMSOL Multiphysics 5.1) as shown in Fig. 3.

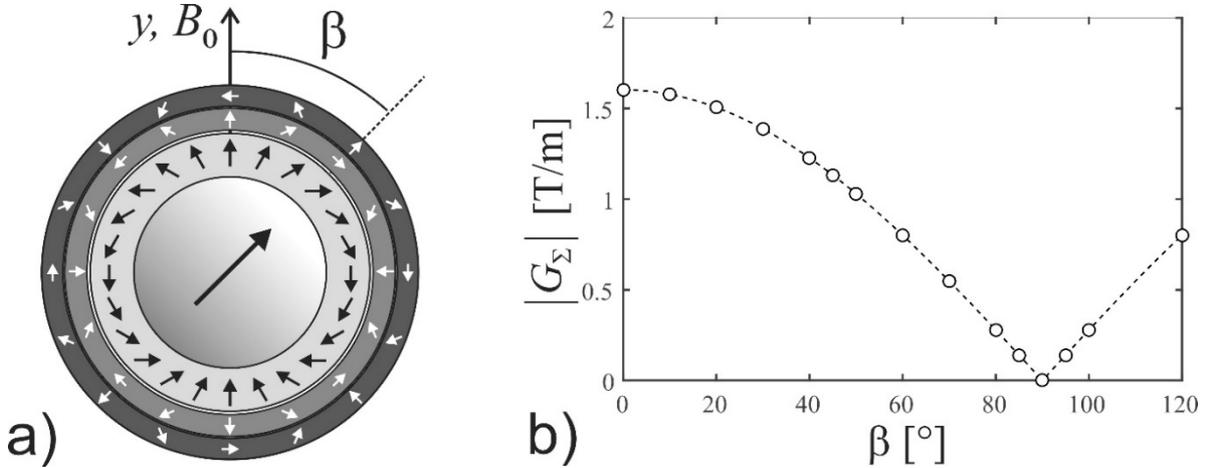

Fig. 3: a) Schematic drawing of a Halbach dipole (light gray, black arrows) concentrically nested in two quadrupoles (darker gray, white arrows). If only one of the quadrupoles (here outer) is turned by an angle $\beta$ the gradient strength changes as shown in b). b) FEM-simulation with $B_R^{Q1} = B_R^{Q2} = 1$ T, $r_i^{Q1} = 15$ cm, $r_o^{Q1} = 16$ cm, $r_i^{Q2} = 16.5$ cm, $r_o^{Q2} = 17.686$ cm. The latter value was chosen so that the simulated values $G_{Q1} = G_{Q2} = 0.801$ T/m although eq. [9] gave $r_o^{Q2} = 17.72$ cm. Both quadrupoles were simulated in 2D as cylinders with 32 segments. The dashed line is a plot of eq. [8]. Note that the direction of $\vec{G}_\Sigma$ is rotated by $\beta$ (and not $2\beta$) relative to $y$, because eq. [8] gives $\measuredangle \vec{G}_\Sigma = \tan^{-1}[\sin 2\beta / (\cos 2\beta + 1)] = \beta$ for $G_{Q1} = G_{Q2}$. In other words, the effective gradient is formed by the vector sum of two vectors of same length and hence its angle is halved.

## Experimental

So far only ideal Halbach systems were considered. However, in reality their continuous magnetization distribution and infinite length has to be discretized and truncated in order to be constructible from permanent magnetic parts. This perturbation of ideality causes inhomogeneities as it can be seen in Fig. 4a/b. When a continuous Halbach magnet (cf. Fig. 4a) is discretized by cutting it into a number of segments with homogeneous magnetization (cf. Fig. 4b) the strongest deviation from the ideal field are observed at the mutual interfaces. However, the production of such cylindrical segments with individual magnetization direction at precise angles is extremely difficult. Therefore, it was suggested [12, 13] to simplify the construction by using magnets with polygonal or circular cross-sections and identical magnetization and mount them such that the magnetization direction fulfills eq. [1] (cf. Fig. 4c). Due to this construction principle production errors of the magnet segments in magnetization direction and strength can be reduced significantly using different approaches [3, 4, 13, 14] resulting in Halbach assemblies with inhomogeneities in the order of ppm.

While all these investigations were done for Halbach dipoles to generate magnetic fields of high homogeneity, Fig. 4 illustrates that the same concept allows to produce discrete Halbach quadrupoles with very homogeneous field gradients. However, there are small differences in the experimental demands for such gradients in MRI. Firstly, they do not need to be as strong as possible. Therefore, the magnet concentration can be reduced, which reduces the gradient amplitude proportional to the area/volume covered by magnetic material (cf. Fig. 4b to 4f). Secondly the homogeneity requirements for the gradient field are less demanding in image reconstruction as nicely demonstrated in [6].

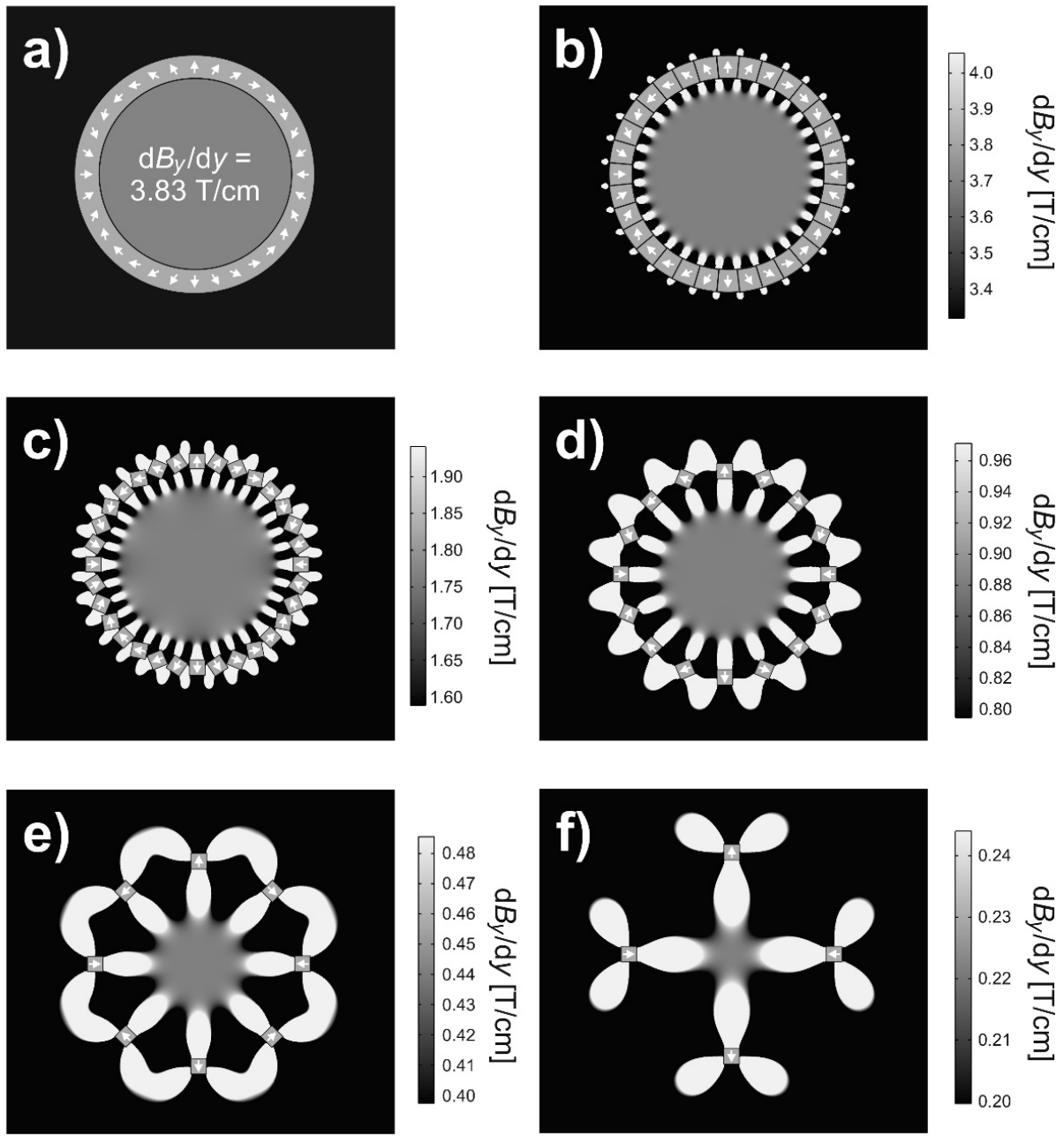

Fig. 4: a) ideal Halbach of magnetic material ($B_R = 1.0$ T, $r_i = 10$ cm, $r_o = 12.37$ cm) with a continuous distributed magnetization direction, producing an homogeneous gradient of 3.83 T/cm as given by eq. [3]. b)-f) 2D FEM-simulation of discrete Halbach quadrupoles. The grayscale images show the produced gradient $dB_y/dy$ of the overlaid magnetic material (gray with white arrows indicating the magnetization direction). The range of the gradient strength is displayed on the right of each image and adjusted to the central gradient value ± 10 %. b) Discretized version of a) consisting of 32 segments producing $G_c = 3.69$ T/cm in the center. c-f) Discrete variations using different numbers of magnets of quadratic cross section with 1.58 cm side length. c) 32 magnets with $G_c = 1.76$ T/cm (corresponding to $f = 47.7$ % of magnet material relative to the amount used in a and b). d) 16 magnets with $G_c = 0.88$ T/cm and $f = 23.9$ %. e) 8 magnets with $G_c = 0.44$ T/cm and $f = 11.9$ % and f) 4 magnets with $G_c = 0.22$ T/cm and $f = 5.97$ %.

Hence, a practical design suggestion for an MRI magnet could be a strong, stationary dipolar Halbach magnet encased by two quadrupole rings which can both be rotated relative to the dipole (cf. Fig. 3a). A rotation of both quadrupoles relative to each other can be used to adjust the gradient strength according to eq. [8]. If the angle between both quadrupoles is then locked and they are both rotated by 90°, full *k*-space can be acquired via projections. If image reconstruction methods are available which tolerate modest gradient inhomogeneities the density of magnets in both quadrupoles can be greatly reduced as shown in Fig. 4. Finally these two ideas can be merged in a magnet design similar to that depicted in Fig. 5a, where four magnets (Q1) are meant to be stationary while four others (Q2) can be rotated on the same perimeter around a dipole. In this way the strength of the gradient can be adjusted (see Fig. 5b) in a very compact design. In this arbitrarily chosen geometry the gradient can be varied from 44.5 to 53.0 mT/m and other values are possible by changing shape and/or strength of Q1 and Q2. While one quadrupole (Q2) is turned to adjust the gradient strength, the other magnets of Q1 can be moved away from their initial position to homogenize the fields [3]. Although the gradient strength can only be changed and not completely cancelled in this device such a design might be sufficient for most MRI applications where the gradient strength merely needs to be adjusted to sample size and spatial resolution.

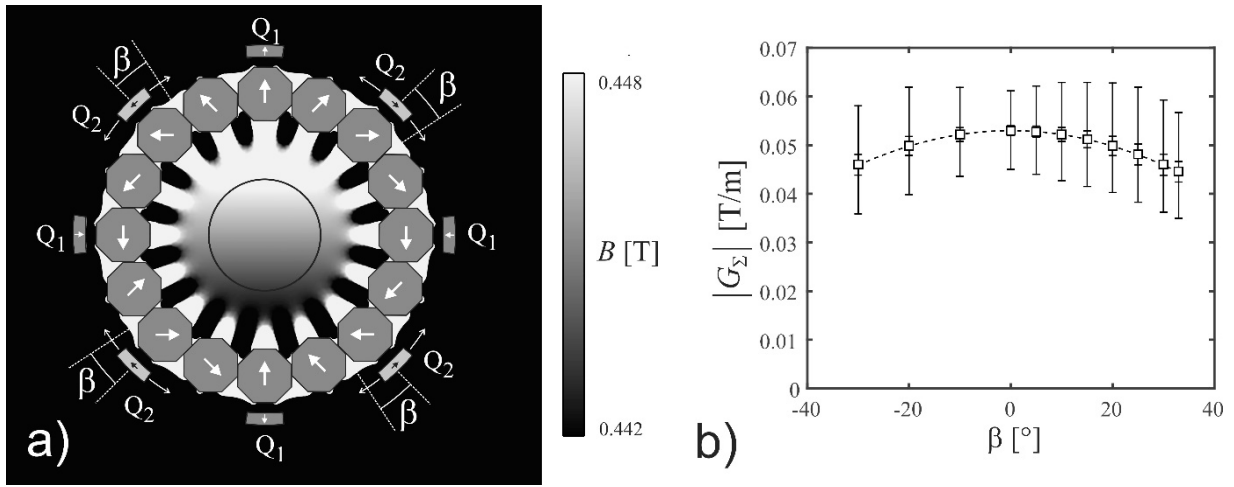

Fig. 5: 2D FEM-simulation of a possible magnet design. a) Design and flux: The inner Halbach dipole consists of 16 octagonal magnets of 3.35 cm in diameter and a remanence of $B_R = 1.5$ T producing a field of 0.45 T over an accessible inner volume of 20 cm in diameter. The quadrupole is made of eight cylindrical segments ($B_R = 0.3$ T, $r_i = 16$ cm, $r_o = 17$ cm) of which four (Q1) are stationary while the others can be turned by an angle β. Each subset of 4 magnets produces a gradient of GQ1 = GQ2 = 0.026 T/cm. The gray scale image shows the magnitude of the generated flux in the range of the gradient field, i.e. 0.44 to 0.448 T. b) Effective gradient $|\vec{G}_\Sigma|$ produced by turning Q2 by an angle β. The inner error-bars represent the standard deviation, the outer the minimal/maximal values in an analyzing circle of 10 cm in diameter (black circle in center of a)). The dashed line is a plot of eq. [8]

## Conclusion

The concept of producing homogeneous magnetic field gradients by superimposing Halbach quadrupoles to Halbach dipoles was introduced together with possibilities to rotate the direction and scale the amplitude of the resulting gradient by turning one or two quadrupoles. The parameters of this design were equated and tested by 2D-FEM simulations. Furthermore, a first practical design was suggested and simulated.

Such homogeneous, constant gradients can be used to make the compact MRI method proposed in [6] more simple, robust and applicable and probably weaken the demands for multiple receive coils and channels, because nonbijective projections of object space to encoding space are reduced or completely avoided.

However, all considerations were only done in 2D and a real MRI system will need 3D resolution. This can either be introduced by shaping the magnetic field in the remaining third dimension such that only a thin slice will be excited [15] or by using an openable Halbach sphere equipped with quadrupoles that can rotate in 3D [4, 16].

It is hoped that these ideas will help to construct simple, robust and effective MRI systems suitable for medical use in third world countries.

## Acknowledgements

The author wants to thank his wife, Friederike Schmid (Physics Dept., University of Mainz), for carefully checking and commenting the manuscript. Furthermore, Lukas Winter (Ultra-High Field Facility, Max Delbrück Center for Molecular Medicine, Berlin-Buch) is acknowledged for stimulating discussions and his enormous engagement to design, build and disseminate an open source MRI system suitable for third world countries.